\documentclass{article}
\usepackage{dcase2024,amsmath,graphicx,url,times,booktabs, tabularx}
\usepackage{multirow}
\usepackage{cite}

\title{WildDESED: An LLM-Powered Dataset for Wild Domestic Environment\\ Sound Event Detection System}

\name{Yang Xiao and Rohan Kumar Das}
\address{Fortemedia Singapore, Singapore\\ \small \{xiaoyang, rohankd\}@fortemedia.com}

\begin{document}

\ninept
\maketitle

\begin{sloppy}

\begin{abstract}
This work aims to advance sound event detection (SED) research by presenting a new large language model (LLM)-powered dataset namely wild domestic environment sound event detection (WildDESED). It is crafted as an extension to the original DESED dataset to reflect diverse acoustic variability and complex noises in home settings. We leveraged LLMs to generate eight different domestic scenarios based on target sound categories of the DESED dataset. Then we enriched the scenarios with a carefully tailored mixture of noises selected from AudioSet and ensured no overlap with target sound. We consider widely popular convolutional neural recurrent network to study WildDESED dataset, which depicts its challenging nature. We then apply curriculum learning by gradually increasing noise complexity to enhance the model's generalization capabilities across various noise levels. Our results with this approach show improvements within the noisy environment, validating the effectiveness on the WildDESED dataset promoting noise-robust SED advancements.
\end{abstract}

\begin{keywords}
sound event detection, DESED, noisy scenario, noise robust SED, curriculum learning
\end{keywords}

\section{Introduction}
\label{sec:intro}
Sounds play a vital role in our lives, helping us understand our surroundings and notice changes. Sound event detection (SED)~\cite{sed,sed2} is essential for interpreting and responding to our environment, with applications ranging from urban noise management to smart-home technologies~\cite{safety1} and security systems~\cite{home}. 
SED has made great strides~\cite{fmsg1,fmsg2,dcase1}, thanks to diverse datasets~\cite{towards} tailored for specific scenarios. Google AudioSet~\cite{audioset} provides a wide array of sounds, and MAVD~\cite{mavd} focuses on traffic noise. Among various SED datasets, DESED~\cite{desed1,desed2} is well known for its focus on domestic environments, which makes it the most utilized dataset for home sound event research. However, DESED faces challenges in comprehensively representing the unpredictable and complex nature of household sounds. Hence, there exists scope for covering a wide range of domestic scenarios with common background noises that can occur in a household.

The quest for noise robustness in SED has led to the development of new methodologies and datasets~\cite{noise4,noise5} aimed towards improving performance under challenging conditions such as noisy urban environments. Innovations by researchers like Neri et al.\cite{noise1}, Serizel et al.\cite{noise3}, and Wan et al.~\cite{noise2} have pushed the boundaries of SED systems by integrating deep learning and audio enhancement techniques. These studies, however, predominantly address controlled or semi-controlled environments, leaving a gap for SED systems to effectively detect sound events in the less predictable, `wild' conditions in domestic environments.

Addressing this gap, our research contributes to the field by introducing a new dataset namely, {\it wild domestic environment sound event detection} (WildDESED). We proposed carefully selecting noise types from the AudioSet that accurately represent real home environments but are distinct from DESED's target sounds. This artificial selection could be challenging because of the bias and unnatural correlations. Large language models (LLMs)~\cite{llm0} such as GPT-4, ChatGPT, and Llama have demonstrated remarkable potential to perform various tasks~\cite{llm1,llm2,llm3} in recent years. In this regard, we utilized the strong capabilities of LLMs to analyze and summarize acoustic data for selecting specific noises. This helped us to design eight different scenarios that blend the noises with target sounds, simulating authentic domestic environments. The noises are divided into four categories based on their sources and acoustic properties, allowing for a diverse and realistic combination with target sounds. This novel approach has culminated in the creation of WildDESED dataset, specifically designed to enhance SED research in dynamic and natural home environments.


Building on this foundation, our research not only introduces the WildDESED dataset, but also explores the application of curriculum learning in the context of SED to tackle the challenges posed by domestic noisy environments. Curriculum learning~\cite{curriculum1,curriculum4,curriculum5} is a training approach that improves models for noisy speech~\cite{curriculum2,curriculum3,ucil} and audio by starting with simpler, less noisy data and gradually increasing the noise level. This method is similar to the way how the humans learn and helps models adjust from clean to noisy sounds more effectively. In this work, we applied curriculum learning to the baseline convolutional recurrent neural network (CRNN)~\cite{crnn,meanteacherbaseline3,fmsg3} model using the WildDESED dataset for our studies. The novelty of this work lies in the proposal of a new {\it in-the-wild} dataset for advancing SED research and exploring curriculum learning as an approach to develop noise-robust SED systems. The WildDESED dataset has been made publicly available\footnote{https://github.com/swagshaw/WildDESED}.

\section{Related work}
The WildDESED dataset is an extension to the original DESED dataset, which is a foundational resource featuring 10 target sound classes pivotal for understanding the sounds in home environments. The DESED dataset consists of the following subsets: The weak set, with 1,578 real recordings labeled with weak annotations, captures the presence of sound classes without temporal specifics. The unlabeled training set includes 14,412 real, unlabeled recordings. The test set comprises of 1,168 real recordings with strong annotations to assess model performance. These three subsets are real-world recordings from AudioSet. The training synth set contains 10,000 synthetic recordings with strong annotations~\cite{impact}, detailing exact temporal boundaries. The synth validation set has 2,500 synthetic recordings with strong annotations for model validation during development. These two synthetic subsets are generated with the Scaper. Their background files are extracted from SINS~\cite{sins}, TUT~\cite{tut}, MUSAN~\cite{musan}, or YouTube and have been selected because they contain a very low amount of our sound event classes. We propose to simulate more diverse and complex noisy scenarios that are not covered by the original DESED dataset and also introduce a controlled variability for testing. 
\section{WildDESED}
\label{sec:method}

We extend DESED to the WildDESED for in-the-wild scenarios for domestic environments by considering three primary set of questions to address as follows: 
\begin{itemize}
    \item What type of background noises do we use?
    \item What are the domestic scenarios we choose?
    \item How do we mix the background noises to the scenarios?
\end{itemize}
\begin{table}[t!]
\centering
\vspace{-2mm}
\caption{A summary of different background noises used in WildDESED dataset.}
\vspace{1mm}
\label{tab:noise-table}
\resizebox{\columnwidth}{!}{%
\begin{tabular}{c|c|c}
\hline
Noise                   & Occurrences & Duration (Second) \\ \hline
Bird chirping outside        & 9,847   & 7,523              \\
Car passing by outside       & 311    & 862               \\
Chair moving                 & 343    & 359               \\
Clock ticking                & 2,5777  & 2,662              \\
Coffee machine               & 6      & 30                \\
Door closing                 & 335    & 196               \\
Fan noise                    & 117    & 958               \\
Footsteps                    & 6,243   & 2,101              \\
Light rain                   & 159    & 1,379              \\
Refrigerator humming         & 58     & 456               \\
TV playing in the background & 805    & 7,191              \\
Wind blowing                 & 5,467   & 48,648             \\ \hline
Total                        & 49,468  & 72,365             \\ \hline
\end{tabular}%
}
\vspace{-4mm}
\end{table}
GPT-4 is an advanced language model that builds on the GPT-3 architecture but uses a larger amount of training data. It includes the latest techniques to enhance understanding of natural language. In the following subsections, we will detail how we leverage GPT-4 to address each of these questions, outlining the methodology behind the creation of the WildDESED dataset. This new dataset aims to bridge the gap between the controlled environment of existing datasets and the dynamic, often unpredictable nature of real-world domestic soundscapes, thus expanding the potential for noise-robust SED research in truly `wild' home scenarios.

\subsection{What type of background noises do we use?}
To construct the WildDESED dataset, we initiated our process with the foundational DESED dataset, which identifies 10 distinct sound events in 10-second audio clips. The events in DESED include diverse household sounds like alarms, appliances, pets, and running water. We input the total 356 classes from the strongly annotated subset of AudioSet to the GPT-4 together with the 10 DESED classes. 
Then we guide GPT-4 by the following prompt: 

\textit{``Select noise classes from the 356 strongly annotated AudioSet classes, alongside the 10 DESED classes ensuring clear delineation and no overlap with DESED's sound events. Further, apply thorough filtering to exclude any AudioSet classes similar to DESED target classes, preserving the distinctiveness of the dataset."}

Considering the output of GPT-4, we enhanced DESED with selected events from the strongly annotated subset of AudioSet, ensuring clear delineation and no overlap with DESED's sound events. A thorough filtering process was applied to exclude any AudioSet classes that are very similar to target classes of DESED dataset, preserving the distinctiveness of our dataset. Table~\ref{tab:noise-table} displays the outcome of our selection process, listing the types and quantities of noise clips integrated into WildDESED. We included a spectrum of sounds both indoor, like the clock ticking, and outdoor, such as birds chirping that capture the essence of a domestic environment. The `clock ticking' class, for instance, has the largest event count, while `wind blowing' spans the greatest duration, together reflecting the continuous and transient nature of home sounds.

This dataset construction ensures WildDESED encompasses a rich and authentic array of domestic noises, ready to challenge and advance SED systems in recognizing the events under complex acoustic home environments.

\subsection{What are the domestic scenarios we choose?}
For the WildDESED dataset, we still have to map the selected 12 noise classes with our 10 target classes. We input them to GPT-4 and use the following prompt:

\textit{``Create eight different domestic scenarios so that they should map 12 selected noise classes to the 10 target classes from the DESED dataset, crafting authentic household soundscapes. Ensure the scenarios reflect typical sounds one would encounter in a household environment."}

Considering the output of LLM, we crafted eight different domestic scenarios, each mapping to target classes from the DESED dataset to create authentic soundscapes one would encounter in a household. These scenarios are constructed to reflect the typical activities and the accompanying sounds in a domestic environment.

\begin{itemize}
    \item \textbf{Morning Routine:} Associated with \textit{`Blender'} target sounds, this scenario captures the essence of the morning with `Light rain', `Refrigerator humming', `Clock ticking', and `TV playing in the background'.
    \item \textbf{Home Office:} Linked to \textit{`Speech'} as the target class, it includes background sounds of `Car passing by', `Fan noise', and `Footsteps', emulating a work-from-home setting.
    \item \textbf{Household Chores:} Representing \textit{`Vacuum cleaner'} noises as the target, this scenario combines `Door closing', `Chair moving', and `Footsteps' as background to depict cleaning activities.
    \item \textbf{Late-night:} Tied to the \textit{`Electric shaver toothbrush'} target sound, offering the `Clock ticking' and `Light rain' as a backdrop for night-time routines.
    \item \textbf{Cooking:} Merging the target sounds of \textit{`Frying'} and \textit{`Dishes'} with `Coffee machine' buzzes and `Refrigerator humming', this scenario is bustling with culinary activity.
    \item \textbf{Pet Care:}  Incorporating target sounds of \textit{`Cat'} and \textit{`Dog'}, this setting is further brought to life with `Bird chirping outside' and `TV playing in the background'.
    \item \textbf{Bathroom Routine:} Linked to \textit{`Running water'} as the target sound, with added `Fan noise' and `Wind blowing', simulating personal care sounds.
    \item \textbf{Emergency:} Associated with the \textit{`Alarm bell ringing'} target sound, it layers urgent sounds like `Refrigerator humming' and `Fan noise' with `Clock ticking' and `Car passing by'.
\end{itemize}
\begin{figure}[t]
  \centering
  \centering
  \includegraphics[width=\linewidth]{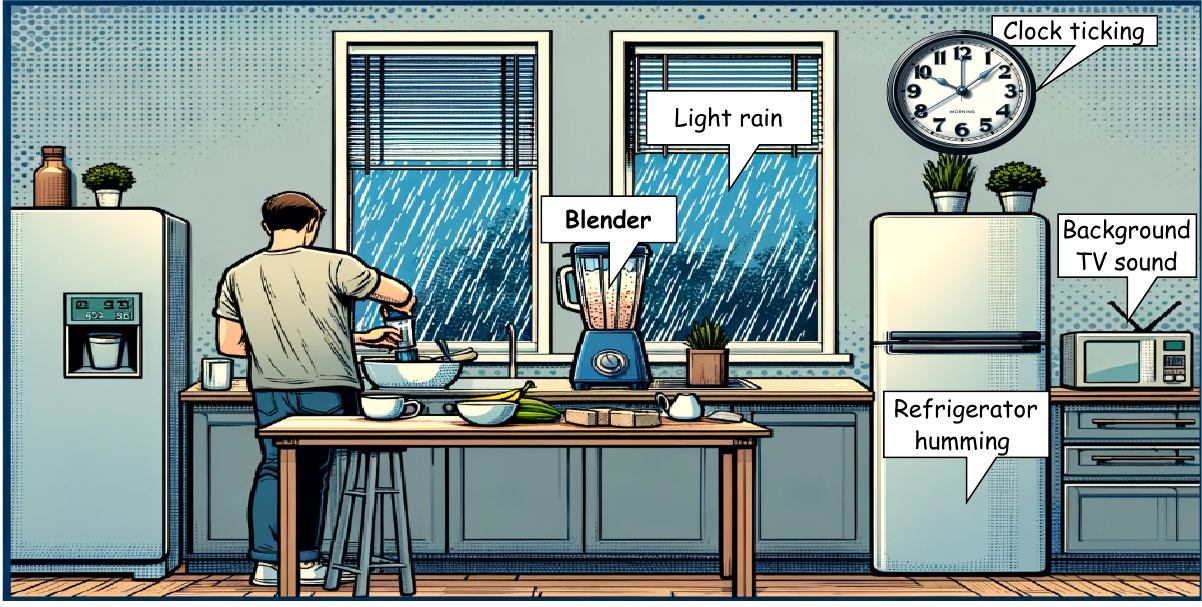}
  \vspace{-6mm}
\caption{Illustration\protect\footnotemark of Morning Routine Scenario out of the total eight scenarios in WildDESED dataset. In the scenario, key target sound events are written in bold fonts, along with added different background noises to simulate real-life settings.}
    \label{fig:scenario}
    \vspace{-4mm}
\end{figure}
Each scenario's sound design is a thoughtful blend of target and noise classes, chosen to challenge the detection capabilities of SED systems within the rich and varied auditory context of a home environment. To illustrate our scenarios, we present a Figure~\ref{fig:scenario} that showcases one typical scenario out of the eight: it is the `Morning Routine Scenario'. This figure highlights the key target sound event `Blender' within this scenario, incorporating strategically placed background noises to simulate the real-life acoustic challenges found in domestic settings. 

\subsection{How do we mix the background noises to the scenarios?}

In the WildDESED dataset, the integration of background noises into the selected domestic scenarios is meticulously structured around a quadrant based on the acoustic characteristics of the noises. The quadrant categorizes noises into four groups: Ambient Environmental Sounds, Human-related and Intermittent Sounds, Mechanical and Electronic Sounds, and Nature and Outdoor Sounds, as illustrated in Figure~\ref{fig:quadrant}.

\begin{itemize}
    \item For \textbf{Ambient Environmental Sounds}, such as `Light rain' and `Wind blowing', we repeated these sounds to cover the entire duration of the audio clip from the original DESED dataset. These sounds are mixed at a low intensity to ensure they provide a consistent background atmosphere without overpowering the primary sound events. The rationale behind this is to create an unobtrusive ambient layer that emulates the continuous presence of these sounds in a typical home environment.
    \item Sounds like `Footsteps', `Door closing', and `Chair moving' fall into the \textbf{Human-Related and Intermittent Sounds} category. These are inserted at random intervals to simulate the sporadic nature of human movement and activities within a home. The volume and frequency of these sounds are varied $\pm 10 \%$ range to reflect the realistic and unpredictable nature of their occurrence in daily life.
    \item \textbf{Mechanical sounds}, including `Clock ticking' and `Coffee machine', are inserted at specific points to coincide with the actions they represent, such as a coffee machine being used during morning routines. The volume is set to be noticeable but not overwhelming, ensuring the sound is recognized as a part of the scenario without becoming a large distraction.
    \item Lastly, \textbf{Nature and Outdoor Sounds } like `Car passing by outside' and `Bird chirping outside' are incorporated randomly to enhance the realism of external environmental influences. The volume may fluctuate to mimic the variable volume of these sounds in real settings, contributing to the unpredictability and diversity of the overall soundscape.
\end{itemize}
\footnotetext{Figures generated using DALL-E-2 (https://openai.com/dall-e-2)}
Each noise type and its corresponding mixing approach are tailored to maintain the authenticity of the domestic scenarios. This methodical and scenario-specific approach to mix noises ensures that the WildDESED dataset not only presents a challenge for SED systems but also closely reflects the complex acoustic environments of actual domestic settings.
\begin{figure}[t]
  \centering
  \vspace{-3.5mm}
\includegraphics[width=0.8\linewidth]{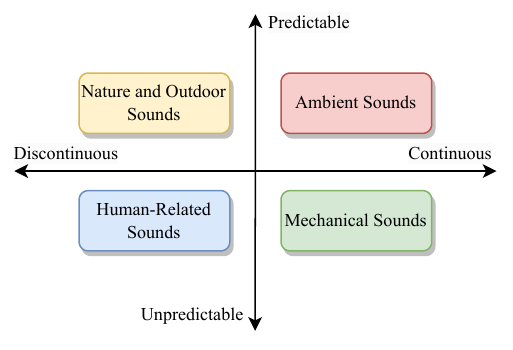}
\vspace{-4mm}
  \caption{Quadrant showing four groups of noise types based on their acoustic characteristics considered in the WildDESED.}
  \label{fig:quadrant}
\end{figure}
\begin{figure}[t]
  \centering
  \includegraphics[width=\linewidth]{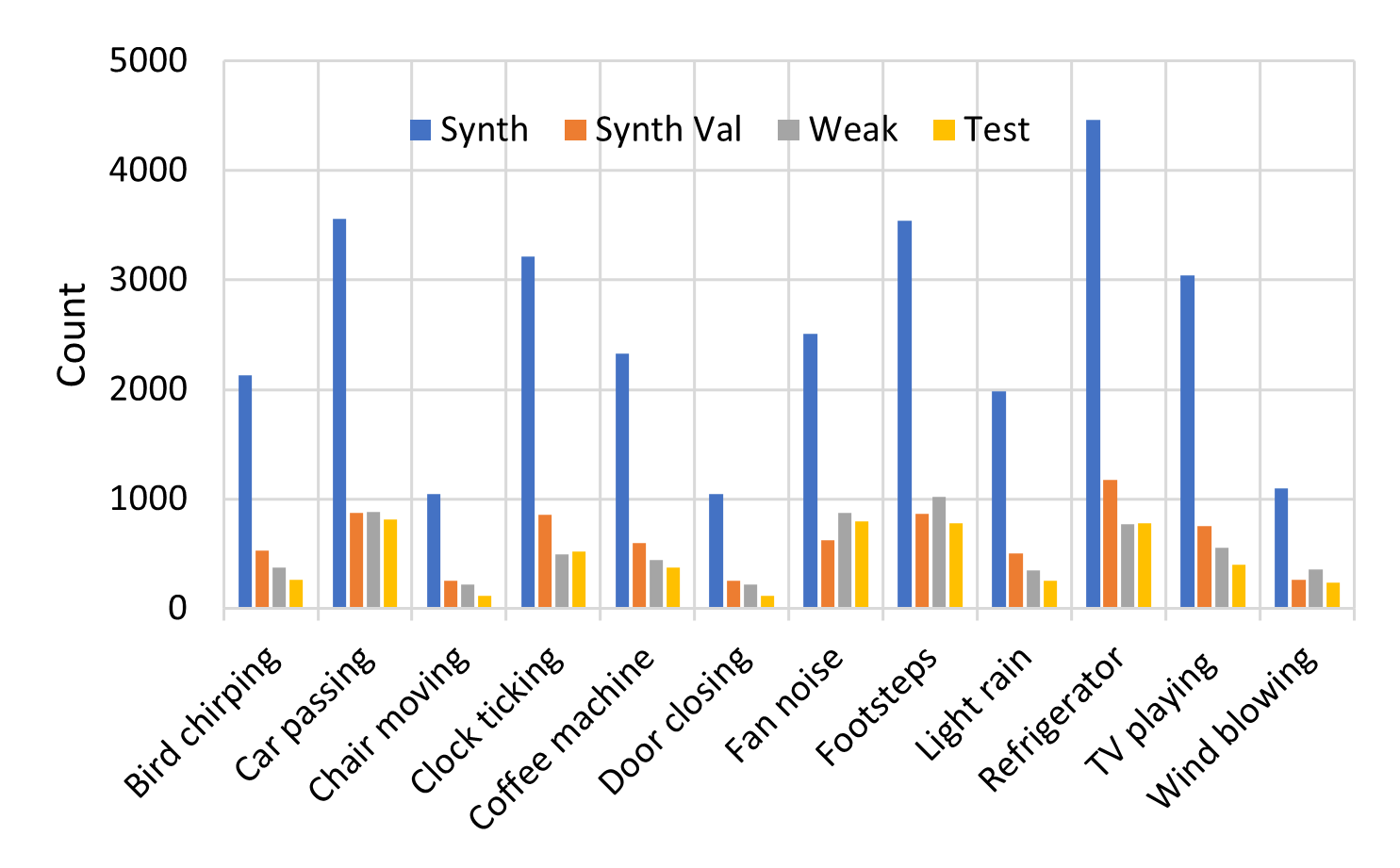}
  \vspace{-4mm}
  \caption{Statistics of noises in the WildDESED subsets.}
  \label{fig:barchart1}
\end{figure}

\begin{table*}[t]
\vspace{-2mm}
\centering
\caption{Performance in PSDS1 (P1), PSDS2 (P2) and PSDS1 + PSDS2 (P1 + P2) of the proposed curriculum learning (CL) approach on the DESED devtest set and our proposed WildDESED (W) dataset with SNR in dB.}
\label{tab:my-table}
\resizebox{\textwidth}{!}{%
\begin{tabular}{|c||ccc||cccccccccccc|}
\hline
\multirow{3}{*}{\textbf{Model}} &
  \multicolumn{3}{c||}{\multirow{2}{*}{\textbf{Performance on DESED}}} &
  \multicolumn{12}{c|}{\textbf{Performance on WildDESED}} \\ \cline{5-16} 
 &
  \multicolumn{3}{c||}{} &
  \multicolumn{3}{c||}{10dB} &
  \multicolumn{3}{c||}{5dB} &
  \multicolumn{3}{c||}{0dB} &
  \multicolumn{3}{c|}{-5dB} \\ \cline{2-16} 
 &
  \multicolumn{1}{c|}{P1} &
  \multicolumn{1}{c|}{P2} &
  P1 + P2 &
  \multicolumn{1}{c|}{P1} &
  \multicolumn{1}{c|}{P2} &
  \multicolumn{1}{c||}{P1 + P2} &
  \multicolumn{1}{c|}{P1} &
  \multicolumn{1}{c|}{P2} &
  \multicolumn{1}{c||}{P1 + P2} &
  \multicolumn{1}{c|}{P1} &
  \multicolumn{1}{c|}{P2} &
  \multicolumn{1}{c||}{P1 + P2} &
  \multicolumn{1}{c|}{P1} &
  \multicolumn{1}{c|}{P2} &
  P1 + P2 \\ \hline\hline
CRNN &
  \multicolumn{1}{c|}{0.344} &
  \multicolumn{1}{c|}{0.543} &
  \textbf{0.887} &
  \multicolumn{1}{c|}{0.222} &
  \multicolumn{1}{c|}{0.409} &
  \multicolumn{1}{c||}{0.631} &
  \multicolumn{1}{c|}{0.148} &
  \multicolumn{1}{c|}{0.302} &
  \multicolumn{1}{c||}{0.450} &
  \multicolumn{1}{c|}{0.064} &
  \multicolumn{1}{c|}{0.174} &
  \multicolumn{1}{c||}{0.238} &
  \multicolumn{1}{c|}{0.017} &
  \multicolumn{1}{c|}{0.078} &
  0.095 \\ \hline
CRNN (W) &
  \multicolumn{1}{c|}{0.200} &
  \multicolumn{1}{c|}{0.329} &
  0.529 &
  \multicolumn{1}{c|}{0.175} &
  \multicolumn{1}{c|}{0.337} &
  \multicolumn{1}{c||}{0.512} &
  \multicolumn{1}{c|}{0.135} &
  \multicolumn{1}{c|}{0.303} &
  \multicolumn{1}{c||}{0.438} &
  \multicolumn{1}{c|}{0.087} &
  \multicolumn{1}{c|}{0.242} &
  \multicolumn{1}{c||}{0.329} &
  \multicolumn{1}{c|}{0.048} &
  \multicolumn{1}{c|}{0.174} &
  0.222 \\ \hline
CRNN (W+ CL) &
  \multicolumn{1}{c|}{0.265} &
  \multicolumn{1}{c|}{0.461} &
  0.726 &
  \multicolumn{1}{c|}{0.212} &
  \multicolumn{1}{c|}{0.443} &
  \multicolumn{1}{c||}{\textbf{0.655}} &
  \multicolumn{1}{c|}{0.175} &
  \multicolumn{1}{c|}{0.390} &
  \multicolumn{1}{c||}{\textbf{0.565}} &
  \multicolumn{1}{c|}{0.114} &
  \multicolumn{1}{c|}{0.317} &
  \multicolumn{1}{c||}{\textbf{0.431}} &
  \multicolumn{1}{c|}{0.049} &
  \multicolumn{1}{c|}{0.211} &
  \textbf{0.260} \\ \hline
\end{tabular}%
}
\vspace{-6mm}
\end{table*}

In finalizing the composition of the WildDESED dataset, special consideration was given to the representation of the `speech' sound class due to its prevalence and significance in domestic environments. For the `Home Office' scenario in synth set and synth val set, we exclusively selected clips that featured the `speech' class in isolation, omitting any clips where `speech' occurred alongside other sound events.

Figure~\ref{fig:barchart1} displays class-wise statistics for different background noises in each subset of the WildDESED dataset, indicating the prevalence of each noise type, within synth, synth val, weak, and test subsets. Figure~\ref{fig:barchart2} shows scenario-wise statistics for the scenarios in the WildDESED dataset, quantifying how frequently each scenario appears in each subset. Through this detailed dataset structure, WildDESED dataset positions itself as a crucial resource for developing and evaluating SED systems, equipping researchers with the means to advance the field of SED in diverse naturalistic home environments.

\section{Curriculum Learning for Noise-robust SED}
\label{sec:curriculum}
We use a curriculum learning~\cite{curriculum1, curriculum3} method to develop noise-robust SED systems. This approach introduces complexity in stages, starting with simple tasks and gradually integrating noise at various signal-to-noise ratios (SNR), aligning with our goal to augment the model's resilience to noise. 

We have five stages in our methodology, each with an increasing level of noise difficulty. Initially, the model learns from clean audio samples. This foundational step is crucial for establishing an understanding of the sound events without the confounding presence of noise. 
We then incrementally introduce noise, decreasing the SNR by 5dB in subsequent stages. Let \(N\) be the total number of training samples. Given \( k \) noise levels \( L = [L_1, L_2, \ldots, L_k] \), the dataset \( D \) is composed as follows:
\vspace{-2mm}
\begin{equation}
    D = \bigcup_{i=1}^{k} \left\{ D_i \right\}, D_i = \frac{N}{k} \text{ samples at noise level } L_i 
    \vspace{-2mm}
\end{equation}

The \(k\) in our experiment here is 5 including the clean DESED, and noise levels 10dB, 5dB, 0dB, and -5dB are considered. The model's progress is meticulously monitored, and a validation metric \(c\) is used to evaluate learning at each epoch. In our approach, the \(c\) is the intersection f1-score. If \(c\) fails to improve for ten consecutive epochs~\cite{early}, the best-performing model state is reloaded, and the training progresses to the next noise level.


\begin{figure}[t]
  \centering
  \includegraphics[width=0.9\linewidth]{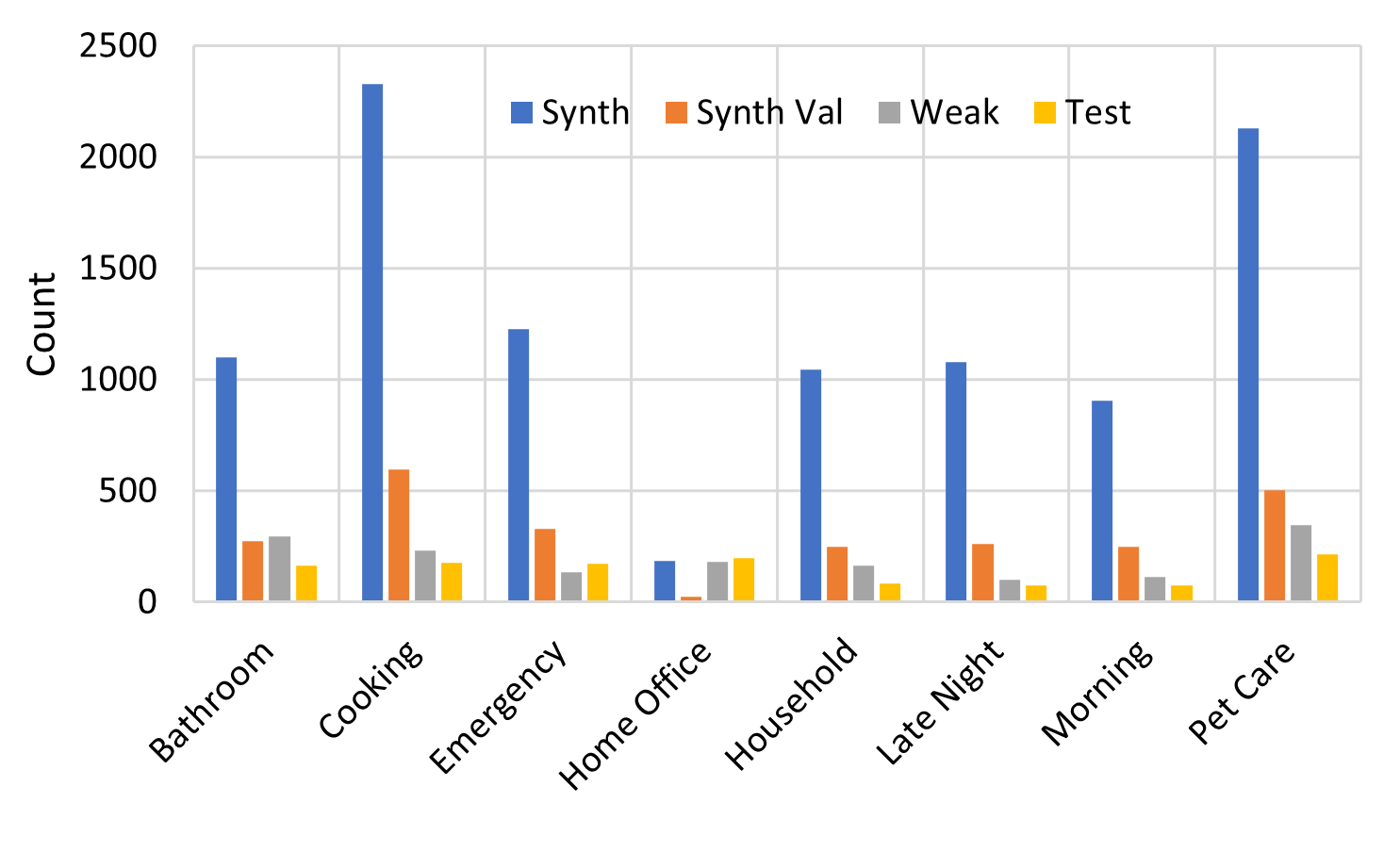}
  \vspace{-8mm}
  \caption{Statistics of the scenarios in the WildDESED subsets.}
  \label{fig:barchart2}
 \vspace{-4mm}
\end{figure}

\section{Experimental Settings}
\label{sec:exp}
\subsection{Dataset and Evaluation Metric}
We considered the DESED dataset and our proposed WildDESED dataset, featuring 10-second audio clips across various subsets. All clips were resampled to 16 kHz mono and segmented using a 2048-sample window and 256-sample hop length for spectrogram extraction and log-mel spectrogram generation. Our systems were evaluated using the threshold-independent polyphonic sound event detection scores (PSDS)~\cite{tpsds} in two scenarios following DCASE 2023 Challenge Task 4A protocol. Scenario-1 focuses on prompt reaction and temporal localization, while Scenario-2 emphasizes on reducing class confusion for SED.

\subsection{Implementation Details}

For our experiments, following the DCASE 2023 Task 4A baseline~\cite{meanteacherbaseline3}, we utilized a batch size of 48 and employed the Adam optimizer with an initial learning rate of 0.001, coupled with an exponential warmup scheduler applied across the first 50 epochs out of a total 200 epochs. To stabilize training, we implemented a mean teacher model with an exponential moving average~\cite{meanteacher} factor set at 0.999. We consider the CRNN~\cite{meanteacherbaseline3} baseline system from DCASE 2023 Task 4A, featuring approximately 1.2 million parameters, ensuring a robust comparison for our curriculum learning approach. 
\section{Results and Discussion}
\label{sec:res}

Table~\ref{tab:my-table} shows the results of our studies on DESED and newly created WildDESED datasets. It is observed that the performance of the baseline CRNN model trained using DESED dataset drops significantly as the noise levels are increased on WildDESED dataset compared to that on the original DESED dataset. We then explore the baseline CRNN model trained using WildDESED data, which we refer to as CRNN (W). We find that CRNN (W) performs better than the original CRNN model when the noise levels on WildDESED are on the higher end (0 dB and -5 dB). However, the performance is comparable for both models when noise level is 5 dB and then the original CRNN model performs better for less noisy scenario of 10dB on WildDESED and on the clean DESED dataset. 

We now focus on the studies for curriculum learning approach applied on the CRNN model trained using WildDESED dataset. We refer this model as CRNN (W+CL) and find that it outperforms both CRNN as well as CRNN (W) models for all noise levels on the WildDESED dataset. This highlights the scope of curriculum learning approach for developing noise-robust SED systems using WildDESED dataset for unseen complex domestic settings. We also note that the CRNN model trained on the clean DESED performs the best on the  DESED test due to the matched conditions. However, the model CRNN (W+CL) with curriculum learning certainly helps to boost the performance of the CRNN (W) model trained on WildDESED dataset to bring it closer that of the CRNN model on DESED test set. The future work will focus on reducing this performance gap on the clean scenario for noise-robust SED models.  
\section{Conclusion}
\label{sec:conclusion}
In this work, we have presented a new dataset referred to as WildDESED to advance SED research under noisy home settings and also explored a preliminary curriculum learning method to develop noise-robust SED systems. We used 12 noises from Audioset to craft the WildDESED dataset considering 8 different scenarios depicting complex home environments by considering assistance from an LLM. The studies conducted showed the scope of curriculum learning approach for developing noise-robust SED systems using the WildDESED dataset. We believe this WildDESED dataset will be useful for future horizons of noise-robust SED research.  

\clearpage

\footnotesize
\bibliographystyle{IEEEtran}
\bibliography{refs}

%
%
%
%
%
%
%
%
%

\end{sloppy}
\end{document}